\documentclass{article}
\usepackage{group21,epsfig}
\title{Icosahedral coincidence rotations}
\author{Johannes Roth\adr{1}  \and\ Reinhard L\"uck\adr{2}}
\address[1]{Institut f\"ur Theoretische und Angewandte Physik,
         Universit\"at Stuttgart,\\
         Pfaffenwaldring 57, D-70550 Stuttgart, Germany}
\address[2]{Max-Planck-Institut f\"ur Metallforschung,\\
         Seestr. 92, D-70174 Stuttgart, Germany}

 \newcommand{\be}{\begin{equation}}
 \newcommand{\ee}{\end{equation}}
 \newcommand{\bea}{\begin{eqnarray}}
 \newcommand{\eea}{\end{eqnarray}}
 \newcommand{\ba}{\begin{array}}
 \newcommand{\ea}{\end{array}}
 \newcommand{\bt}{\begin{tabular}}
 \newcommand{\et}{\end{tabular}}
 \newcommand{\bd}{\begin{displaymath}}
 \newcommand{\ed}{\end{displaymath}}
 
 \newcommand{\bm}[1]{\mbox{\boldmath $#1$}}

 \newcommand{\zr}[1]{\mbox{\hspace*{#1em}}}

 \newcommand{\QQ}{\mbox{{\sf Q}\zr{-0.53}\rule[0.1ex]{0.04em}{1.4ex}\zr{0.53}}}
  
 \newcommand{\RR}{\mbox{\zr{0.1}\rule{0.04em}{1.6ex}\zr{-0.05}{\sf R}}}
 
 \newcommand{\ZZ}{\mbox{\sf Z\zr{-0.45}Z}}
 
 \newcommand{\II}{\mbox{\rm I\zr{-0.23}I}}

\begin{document}
\maketitle
\begin{abstract}
 The coincidence problem for three-dimensional discrete structures
with icosahedral symmetry is reinvestigated. 
We present a parametric description of the
coincidence rotations based on special quaternions, called icosian
numbers. In particular, we give a characterization of the possible
coincidence indices ($\Sigma$-factors) and present a complete list of
the possible rotations with $\Sigma \leq 100$.
\end{abstract}

\section{Introduction}
The symmetries of crystallographic lattices are well known: these are the
operations that map the whole lattice onto itself, so that each
lattice point coincides exactly onto another one. The symmetry
operations are translations, rotations, reflections and combinations
thereof. It is, however, also possible to map only a part of the
lattice points onto the others: a sublattice emerges. If the
sublattice has full rank (i.e. same dimension as the original
lattice), but possibly lower symmetry, it is called a coincidence site
lattice, or CSL for short. The CSL is characterized by its index in
the original lattice. This number is also called the $\Sigma$-factor. 
CSLs are useful for several aspects. 

First of all, there are
mathematical reasons: the point symmetry group of the lattice $\cal G$ is a
subgroup of the symmetry group of the isotropic Euclidean
space, $SO(n, \RR)$. The set of coincidence rotations also forms a group,
and includes the group of symmetries. In particular, 
it contains rotations with  rational angle. Therefore we have $SO(n, \ZZ) 
\subset {\rm CSL-group} \subset SO(n, \RR)$, 
and it would be desirable to know more about the structure
of the CSL group. Since the point symmetry group may be a maximal
finite subgroup of $SO(n, \RR)$, the CSL group must have infinite order.

Second there are crystallographic reasons: A real crystal consists of
several grains. It has been shown that these components are
mostly not oriented arbitrarily with respect to each other, but that two
adjacent grains often have a common coincidence lattice with low
index in the full lattice.
The CSL are important in crystallography because they allow a
nontrivial classification of grain boundaries and because
small-unit-cell CSL grain boundaries seem to be energetically
favoured\cite{sb}. 
It has been shown that coincidences of vertices appear also in
quasicrystalline tilings\cite{sr,dw}.
Quasicrystals also have grain boundaries like ordinary crystals, and
one should therefore know the coincidence site
quasilattices\cite{dw}. Twinning has been observed in
experiment\cite{du,dw}, as well as interfaces between icosahedral and
decagonal quasicrystals\cite{fwu}. Similar to the crystal the CSL
grain boundary should be energetically favoured. To understand
the occuring coincidence indices one has to enumerated the possible CSL.
The grain boundary
between crystals itself may be quasiperiodic as predicted by Rivier and
Lawrence\cite{rl}. Indirect evidence was provided already in the
1970s\cite{stb,tsb}, and later observed in the growth of
quasicrystalline grain at the grain boundary between two
crystals\cite{csp,sp}. 

Some microstructures with icosahedral (pseudo) symmetry are composed
of 'approximants' by twinning and multiple twinning. The grain
boundaries of these nanodomains will also be low energy grain
boundaries, but their orientation relation is different from those
ones described in the present paper.

The first one who has dealt with generating functions and the
enumeration of CSL seems to have been Ranganathan\cite{B0}.
The classification of the crystallographic CSL has been obtained by
several authors in the '70's and '80's\cite{W,N,B,C}, and there exist
tables of cubic and hexagonal CSL. But there has been no systematic
mathematical treatment of the problem until recently, see \cite{B96}
and references therein.  

The interest in CSL was renewed with the discovery of the
quasicrystals. It has been found that this new type of materials also
exhibits multiple grains, twin relationships and coincidence
(quasi-)lattices.

In a quasicrystal the number of generating basis vectors is larger than
the space dimension. This means that there exists no lattice with a
minimal distance in 
physical space but a dense module. The CSL has to be replaced by a
coincidence site module (CSM). A quasilattice is generated if only
certain points are selected from the module by a window function.

Recently the complete classification of all possible CSL/CSM for crystals
and quasicrystals with rotational symmetry of any order and an
analysis of the structure of the CSL/CSM group has been
achieved by Pleasants et al.~\cite{D} with the help of
number theoretical methods. This means that the CSL/CSM case is solved
completely in two dimensions. In three and four dimensions, similar
(but less complete) results are known through the use of quaternions,
see Baake and Pleasants \cite{BP} and Baake \cite{B96}.

In three dimensions the problem has also been adressed by Warrington and
L\"uck\cite{E}, who enumerated the icosahedral CSM with small index,  and by
Radulescu and Warrington\cite{F}. In this paper we will generalize the
systematic and appealing treatment of the crystallographic CSL in
three dimensions by Grimmer\cite{B,C} to icosahedral
quasicrystals. The method is suitable for an implementation in a
computer program and allows an easy enumeration of all CSM up to high
indices. Our main purpose is therefore the presentation of complete
tables of icosahedral CSM representatives.

\section{Formulation of the problem}
A coincindence rotation $C$ maps a lattice $L$ onto a lattice
$L^{\prime}$:
\begin{equation}\label{firsteq}
L^{\prime}=C L 
\end{equation}
The rotation $C$ is not unique, since we can apply symmetry
rotations $R$ and $S$ to the lattices $L = R \tilde{L}$ and $L' = S
\tilde{L}'$, resulting in a 
conjugation of $C^{\prime}$: 
\begin{equation}
S \tilde{L}^{\prime} = C R \tilde{L} \quad {\rm or} \quad
\tilde{L}^{\prime} = S^{-1} C R \tilde{L} 
\end{equation}
By comparison with Eq.\ \ref{firsteq} we may introduce a new
coincidence rotation $C'$:
\begin{equation}
C^{\prime} = S^{-1} C R
\end{equation}
This means that we have to count the coincindence rotations $C$ with
respect to double coset classes:
\begin{displaymath}
(S,R) \in  SO(3) \times SO(3) \cong SO(4)/Z_2
\end{displaymath}
First of all we need a proper parametrization of the rotation
matrices. A three-dimensional matrix has nine entries, but only three are
independent: the axis (of length one) and the angle.  Cayley's parametrization
through quaternions is useful here \cite{B96,BP,B}. 
If ${\bf q}=(\kappa, \lambda, \mu, \nu)$ is the quaternion, the first entry
$\kappa$ parametrizes the angle:
\begin{displaymath}
\cos(\phi) =
\frac{\kappa^2-\lambda^2-\mu^2-\nu^2}{\kappa^2+\lambda^2+\mu^2+\nu^2}
\end{displaymath}
and the vector $(\lambda, \mu, \nu)$ represents the rotation axis.
The rotation matrix is therefore given by:
\begin{displaymath}
R \; = \; 
\frac{1}{|\bm{q}|^2}
\left( 
\begin{array}{ccc}
  \kappa^2 + \lambda^2 - \mu^2 - \nu^2 & 
  -2\kappa\nu + 2\lambda\mu & 2\kappa\mu + 2\lambda\nu \\
  2\kappa\nu + 2\lambda\mu & \kappa^2 - \lambda^2 + \mu^2 - \nu^2 &
  -2\kappa\lambda + 2\mu\nu \\
  -2\kappa\mu + 2\lambda\nu & 2\kappa\lambda + 2\mu\nu &
  \kappa^2 - \lambda^2 - \mu^2 + \nu^2
\end{array} 
\right) 
\end{displaymath}

The norm  of a quaternion is:
\begin{displaymath}
|{\bf q}|^2 = \kappa^2+\lambda^2+\mu^2+\nu^2
\end{displaymath}
We have a two-to-one homomorphism of the quaternions and the rotation
matrices since $R({\bf q})=R(-{\bf q})$. In the quaternionic
formulation of the coincidence problem we get\cite{G} (small letters
indicate quaternions):
\begin{displaymath}
L = C^{\prime} L \quad \rightarrow \quad {\bf l}^{\prime} = {\bf
  c}^{\prime} {\bf l} {\bf c}^{\prime -1}
\end{displaymath}
\begin{equation}
C^{\prime} = S^{-1} C R \quad \rightarrow \quad {\bf c}^{\prime} =
{\bf s}^{-1} {\bf c}{\bf r} \label{crq}
\end{equation}
Similar to the three-dimensional Cayley parametrization there exists
also a four-dimensional one: Due to the homomorphisms between the
quaternions, $SU(2)$ and $SO(3)$ and the homomorphism between $SO(3)
\times SO(3)$, $SU(2) \times SU(2) $ and $SO(4)$, we can parametrize
four-dimensional rotation matrices by a pair of
quaternions\cite{B}. The equation \ref{crq} is thus further transformed
into 
\begin{equation}
\label{poleq}
{\bf c}^{\prime}  =  {\bf s}^{-1} {\bf c}{\bf r} \quad \rightarrow
\quad C^{\prime} = M C 
\end{equation}
Now C represents a four-dimensional vector and M a four-dimensional
rotation matrix parametrized by two quaternions ${\bf r} = (k, l, m,
n,)$ and ${\bf s} = (a, b, c, d)$. The rotation matix has the form: 
\begin{displaymath}
\hspace*{-0.5cm} M \; = \; 
\frac{1}{|\bm{s}||\bm{r}|}
\left( 
\begin{array}{clccc}
 ak+bl+cm+dn & -al+bk+cn-dm & -am-bn+ck+dl & -an+bm-cl+dk\\
 al-bk+cn-dm &  ak+bl-cm-dn & -an+bm+cl-dk &  am+bn+ck+dl\\
 am-bn-ck+dl &  an+bm+cl+dk & ak-bl+cm-dn & -al-bk+cn+dm\\
 an+bm-cl-dk & -am+bn-ck+dl & al+bk+cn+dm &  ak-bl-cm+dn
\end{array} 
\right) 
\end{displaymath} 

\begin{figure}
\caption{The asymmetric unit of the four-dimensional polytope
  \{5,3,3\}. The dash-dotted line is the 5-fold axis, the long-dashed
  lines are 2-fold axis and the short-dashed lines 3-fold axis.\label{asico}}
\centerline{\epsfig{figure=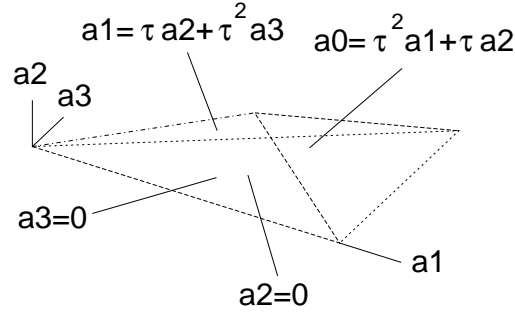,width=5cm,angle=270}}
\end{figure}

What is the meaning of this transformation? Equation \ref{poleq} is nothing
else than the symmetry description of a four-dimensional polytope! A
detailed description of the polytope may be found in Ref.\ \cite{E1}. We 
have transformed the problem of enumerating double coset classes into
the problem of analyzing a four-dimensional polytope. 
If we calculate
the orbit of a certain point of the polytope we find that 
the size of the orbit gives the number of equivalent CSM, and the type of
the orbit tells us the symmetry of the CSM. Practically, we have to
find a standard representation for the CSM. This is the
disorientation\cite{B}: within a class of equivalent coincidence
rotations there is one where the rotation angle transforming $L$ into
$L^{\prime}$ is minimal. Therefore $\kappa$ in ${\bf q}$ is 
maximal. Translated into the four-dimensional language of the polytopes
we get a set of linear inequalities\cite{B} which define an asymmetric
unit. Fig.~\ref{asico} displays the icosahedral case. The
crystallographic cases can be found in Ref.~\cite{C}.

In order to count the orbits we have to fill the polytope with
points. In the cubic case it has been shown\cite{B} that it is
sufficient that the entries of the quaternion ${\bf q}=(\kappa, \lambda,
\mu, \nu)$ are integers. In the icosahedral case we have to use
numbers of the type $m+n\tau$ with $m$ and $n$ integers and $\tau$ the
golden mean $(\sqrt{5}+1)/2$\cite{A}. These
numbers are elements of the ring of icosians generated by the quaternions
$(1,0,0,0)$, $1/2(1,1,1,1)$ and $1/2(\tau,1,-1/\tau,0)$. These quaternions
generate the group $\hat{Y}$ of order 120, and the ring of icosians $\II$
are all integer linear combinations.

There exists a further interesting module: It consists of all points
in $\ZZ[\tau]$, has cubic symmetry and contains the icosahedral
modules as subsets (for details see \cite{B96}). 
The coincidence rotations are generated by the group
$SO(3,\QQ(\tau))$, and the quaternions have entries of type  $m+n\tau$
without any restriction.

\section{Technicalities}
To avoid counting equivalent quaternions several times we have to
demand that three conditions are fulfilled: First, we
take only quaternions whose components have no common factor:
${\rm gcd}(\kappa, \lambda, \mu, \nu) = 1$. Second, the entries of ${\bf q}$
are ordered with decreasing size. Third, the norm of a number
$m+n\tau$\cite{A} is: 
\begin{displaymath}
N(m+n\tau) = m^2 +mn - n^2
\end{displaymath}
This means that
\begin{displaymath}
N(\pm\tau^k) =(-1)^k
\end{displaymath}
with $k$ an integer. To take this last condition into account it is
convenient to restrict the smallest non-zero quaternion component between 1
and $\tau$. 

The index for an icosahedral CSM is given by:
\begin{displaymath}
\Sigma = N (|q|^2) / 2^{\ell}
\end{displaymath}
where $\ell=0$ if one of the (coprime) quaternion components is odd and $\ell=2$ if
all are even.
For the cubic case it is:
\begin{displaymath}
\Sigma = |N({\rm den}(R))|
\end{displaymath}
where ${\rm den}(R)$ is the denominator of the rotation matrix.

\section{Results}
It is easy to count icosahedral CSM and the CSM of $\ZZ[\tau]$ on a
workstation by looping through the unit icosians or through the
quaternions with $m+n\tau$ entires using the algorithm described
above. The correctness 
of the results can be checked by the generating function for the
number of CSM $\#(\Sigma)$ given in
Ref. \cite{B96,A}. We have tabulated the CSM with a rotation axis parallel
to a symmetry axis up to $\Sigma=5000$, with an axis in a general up
to $\Sigma=500$. Some special cases of the order $\Sigma=20000$ have
also been checked. Summary tables of the results are given in
Tab.~\ref{tab1a},\ref{tab1b}, representatives for the icosahedral
case in Tabs.~\ref{tab2a},\ref{tab2b},\ref{tab2c}. 
The cubic case is presented in
Tabs.~\ref{tab1a},\ref{tab1b},\ref{tab1c}, and
\ref{tab3a},\ref{tab3b},\ref{tab3c},\ref{tab3d}. 

It is interesting to note that the number of CSM for the cubic and
icosahedral case are identical if $\Sigma$ is odd, but if $\Sigma$ is
even, $N_{cub}(\Sigma) = 8 * N_{ico}(\Sigma/4)$. 

\section*{Miscellany}
The results described in this paper have been presented at the
XXI. Int. Coll. on Group Theoret. Methods in Physics, July 15th-19th,
in Goslar, Germany. The tables listed in this paper may be obtained
from the authors through email from
 {\sf johannes.roth@itap.physik.uni-stuttgart.de.} 

\section*{Acknowledgements}
We thank Michael Baake for his cooperation during the preparation
of this manuscript.


\begin{table}
\caption{Summary of the CSM orbits.~I.
The part left of the double line are the results for the icosahedral
case, the part to the right is for the cubic module $\ZZ[\tau]^{3}$.
Cn denotes a CSM axis parallel to n-fold axis, M a rotation axis in a
mirror plane and G a rotation axis in a general position. The numbers
$\#(\sigma)$ give the sum of the lengths of the orbits. Entries of type
a*b denote a*b CSM with b algebraic conjugates. The symmetry groups
for $\Sigma < 10$ are explicitly given. Primes indicate a CSM occuring
in a standard cubic lattice.
\label{tab1a}}
\vspace{0.5cm}

\begin{center}
\begin{tabular} {r|r|c|c|c|c|l||c|c|c|c|c|l}
$\Sigma$ & $\#(\Sigma)$ & C5:12 & C3:20 & C2:30 & M:60 & & 
C4:6 & C3:8 & C2:12 & M:24 & G:48 & \\
\hline
  1 &   1 &  & & &  & Y  &   &   &   &   &   & O\\
  4 &   5 &  & & &  & T  &   &   &   &   &   &  \\
  5 &   6 &  & & &  & D5 & 1 &   &   &   &   &  \\
  9 &  10 &  & & &  & D3 & 1 &   &   &   &   & D3' \\
\hline
 11 &  24 & 1*2  &     &     &   &  &   &   & 2 &   & &\\        
 16 &  20 &      &1*1  &     &   &  &   &   &   &   & & \\      
 19 &  40 &      &1*2  &     &   &  &   &   & 2 &   & & \\          
 20 &  30 &      &     &1*1  &   &  &   &   &   &   & & \\  
 25 &  30 &      &     &1*1  &   &  & 1' &   &   & 1 & & \\
 29 &  60 &      &     &1*2  &   &  & 2 &   &   & 2 & & \\
 31 &  64 & 1*2  &1*2  &     &   &  &   &   &   & 2 & & \\
 36 &  50 &      &1*1  &1*1  &   &  &   &   &   &   & & \\  
 41 &  84 & 1*2  &     &1*2  &   &  & 2 &   & 2 & 2 & & \\
 44 & 120 &      &     &     &1*2&  &   &   &   &   & & \\
 45 &  60 &      &     &1*2  &   &  & 2 &   &   &   & 1 & \\
 49 &  50 &      &1*1  &1*1  &   &  &   & 1' &  1 & 1 & & \\
 55 & 144 & 1*2  &     &     &1*2&  &   &   &   & 6 & & \\
 59 & 120 &      &     &     &1*2&  &   &   & 2 & 4 & & \\
 61 & 124 & 1*2  &1*2  &1*2  &   &  & 2 & 2 &   & 4 & & \\
 64 &  80 &      &1*1  &     &1*1&  &   &   &   &   & &  \\
 71 & 144 & 1*2  &     &     &1*2&  &   &   &   & 6 & & \\
 76 & 200 &      &2*2  &     &1*2&  &   &   &   &   & &  \\
 79 & 160 &      &1*2  &     &1*2&  &   & 2 &   & 6 & & \\
 80 & 120 &      &     &     &1*2&  &   &   &   &   & &  \\
 81 &  90 &      &     &1*1  &1*1&  & 1 &   & 1'  & 1 & 1 & \\
 89 & 180 &      &     &1*2  &1*2&  & 2 &   & 2  & 2 & 2 & \\
 95 & 240 &      &     &     &2*2&  &   &   &   & 6 & 2 & \\
 99 & 240 &      &     &     &2*2&  &   &   &   & 4 & 2 & \\
100 & 150 &      &     &1*1  &1*2&  &   &   &   &   & &  \\
101 & 204 & 1*2  &     &1*2  &1*2&  & 2 &   &   & 1 & 3 & \\
109 & 220 &      &1*2  &1*2  &1*2 &  & 2 & 1 &   & 8 & & \\
116 & 300 &      &     &1*2  &2*2  &  &   &   &   &   & &  \\          
\end{tabular}
\end{center}
\end{table}

\begin{table}
\caption{Summary of the CSM orbits.~II. Notation as in
  Tab.~\ref{tab1a}. \label{tab1b}}  
\vspace{0.5cm}
\begin{center}
\begin{tabular}{r|r|c|c|c|c|c||c|c|c|c|c}
$\Sigma$ & $\#(\Sigma)$ & C5:12 & C3:20 & C2:30 & M:60 & G:120 & 
C4:6 & C3:8 & C2:12 & M:24 & G:48 \\
\hline
121 & 408 & 2*2  &     &     &2*(2+1)&         &   &  & 4 & 11 & 2 \\        
124 & 320 &      &2*2  &     &   &      1*2    &   &   &  &  &    \\
125 & 150 &      &     &1*1  &1*2&             & 1 &   &  &  4& 1 \\
131 & 264 & 1*2  &     &     &2*2&             &   &   & 2 & 6 & 2 \\
139 & 280 &      &1*2  &     &2*2&             &   & 2 & 2 & 6 & 2 \\
144 & 200 &      &1*1  &     &1*1&      1*1    &   &   &   &  &   \\
145 & 360 &      &     &2*2  &2*2&             & 4 &   &   & 10 & 2 \\
149 & 300 &      &     &1*2  &2*2&             & 2 &   &   &  8& 2 \\
151 & 304 & 1*2  &1*2  &     &2*2&             &   & 2 &   &  8& 2 \\
155 & 384 & 1*2  &     &     &1*2&      1*2    &   &   &   &  8& 4 \\
164 & 420 &      &     &1*2  &1*2&      1*2    &   &   &   &  &   \\
169 & 170 &      &1*1  &1*1  &2*1&             & 1' & 1' & 1 &2  & 6 \\
171 & 400 &      &1*2  &     &3*2&             &   & 2 & 4 &  2& 6 \\
176 & 480 &      &     &     &4*2&             &   &   &   &  &   \\
179 & 360 &      &     &     &3*2&             &   &   & 2 &  6& 4 \\
180 & 300 &      &     &1*2  &   &      1*2    &   &   &   &  &   \\
181 & 364 & 1*2  &1*2  &1*2  &2*2&             & 2 & 2 &   &  6& 4 \\
191 & 384 & 1*2  &     &     &3*2&             &   &   &   &  8& 4 \\
196 & 250 &      &1*2  &1*1  &1*1&      1*1    &   &   &   &  &   \\
199 & 400 &      &1*2  &     &1*2&      1*2    &   &   &   &  12& 2 \\
\hline
205 & 504 & 1*2  &     &2*2  &1*2&      1*2   & & & & & \\
209 & 960 &      &     &     &6*2&      1*2   & & & & & \\
211 & 424 & 1*2  &1*2  &     &3*2&            & & & & & \\
220 & 720 &      &     &     &4*2&      1*2   & & & & &  \\
225 & 300 &      &     &1*2  &1*2&      1*1   & & & & & \\
229 & 460 &      &1*2  &1*2  &1*2&      1*2   & & & & & \\
236 & 600 &      &     &     &5*2&             & & & & &  \\
239 & 480 &      &     &     &2*2&      1*2   & & & & & \\
241 & 484 & 1*2  &1*2  &1*2  &3*2    &         & & & & & \\
244 & 620 &      &2*2  &1*2  &       &      2*2   & & & & &  \\
245 & 300 &      &     &1*2  &       &      1*2   & & & & & \\
251 & 504 & 1*2  &     &     &2*2    &      1*2   & & & & & \\
256 & 320 &      &1*1  &     &1*(2+1)& 1*1  & &  & & &  \\
\end{tabular}
\end{center}
\end{table}

\begin{table}
\caption{Summary of the CSM orbits.~III. This table lists the
coincidence rotations for the cubic module $\ZZ[\tau]^{3}$ with even
$\Sigma$. Notation as in Tab.~\ref{tab1a}.\label{tab1c}}
\vspace{0.5cm}
\begin{center}
\begin{tabular}{ r | r | r | c | c }
$\Sigma$ & $\#(\Sigma)$ & C3:8 & M:24 & G:48 \\
\hline
       4 & 8     &  1 &                 &   \\
      16 & 32    &  1  &            1   &    \\
      20 & 48    &     &           2  &     \\
      36 & 80    &  1    &          1  &     1 \\
      44 & 192   &       &         8&       \\
      64 & 128   &  1      &        3       &1\\
      76 & 320   &  4       &       4      & 4\\
      80 & 192   &          &      4     &  2\\
     100 & 240   &           &     4    &   3\\
     116 & 480   &            &    8   &    6\\
     124 & 512   &  4           &  12  &     4\\
     144 & 320   &  1            &  7 &      3\\
     164 & 672   &               &12&       8\\
     176 & 768   &   &             8      &12\\
     180 & 480   &    &            8     &  6\\
     196 & 400   &  2   &           4    &   6\\
     220 & 1152  &      &         16   &   16\\
     236 & 960   &       &         8  &    16\\
     244 & 992   &  4      &       16  &    12 \\
     256 & 512   &  1       &       7&       7\\

\end{tabular}
\end{center}
\end{table}

\begin{table}
\caption{Representatives for coincidence rotations with $\Sigma <
100$ for the icosahedral module.~I. The two sections indicate rotation
axis along 5- and  3-fold 
symmetry axis, respectively. The coordinates are given as
$m+\tau n$. The first column lists the conjugate pairs and the last
column the disorientation angle.
\label{tab2a}}
\vspace{0.5cm}
\begin{center}
\begin{tabular}{c|c|c|c|c|c|c}
conjugates & $\kappa$ & $\lambda$ & $\mu$ & $\nu$ & $\Sigma$ & angle \\
\hline 
  a  a & 0    1 &  1   -1 &  0    0 &  1    2 &    5 &      36.000000\\[0.5ex]

  a  b & 0    1 &  1   -1 &  0    0 &  3    2 &   11 &    19.464600\\
  b  a & 0    1 &  1   -1 &  0    0 &  3    0 &   11 &    27.227642\\[0.5ex]

  a  b & 0    1 &  1   -1 &  0    0 &  5    2 &   31 &    13.290735\\
  b  a & 0    1 &  1   -1 &  0    0 &  1    4 &   31 &    23.637167\\[0.5ex]

  a  b & 0    1 &  1   -1 &  0    0 &  5    4 &   41 &    11.107199\\
  b  a & 0    1 &  1   -1 &  0    0 &  3    4 &   41 &    15.125998\\[0.5ex]

  a  b & 0    1 &  1   -1 &  0    0 &  7    4 &   55 &     8.772344\\
  b  a & 0    1 &  1   -1 &  0    0 &  5    0 &   55 &    16.535397\\[0.5ex]

  a  b & 0    1 &  1   -1 &  0    0 & -1    6 &   61 &    30.034790\\
  b  a & 0    1 &  1   -1 &  0    0 &  5   -4 &   61 &    32.070736\\[0.5ex]

  a  b & 0    1 &  1   -1 &  0    0 &  5   -2 &   71 &    21.850590\\
  b  a & 0    1 &  1   -1 &  0    0 &  2    1 &   71 &    31.020212\\[0.5ex]
\hline 
  a  a & 0    1 &  1   -1 &  0    0 &  1    1 &    4 &    44.477505\\[0.5ex]

  a  a & 0    1 &  1   -1 &  0    0 &  3   -3 &    9 &    60.000000\\[0.5ex]

  a  a & 0    1 &  1   -1 &  0    0 &  3   -1 &   16 &    31.044971\\[0.5ex]

  a  b & 0    1 &  1   -1 &  0    0 &  3    1 &   19 &    20.724964\\
  b  a & 0    1 &  1   -1 &  0    0 &  1    3 &   19 &    26.101496\\[0.5ex]

  a  b & 0    1 &  1   -1 &  0    0 & -1    5 &   31 &    35.127438\\
  b  a & 0    1 &  1   -1 &  0    0 & -3    7 &   31 &    53.023972\\[0.5ex]

  a  a & 0    1 &  1   -1 &  0    0 &  3    3 &   36 &    15.522505\\[0.5ex]

  a  a & 0    1 &  1   -1 &  0    0 &  5   -5 &   49 &    38.213223\\[0.5ex]

  a  b & 0    1 &  1   -1 &  0    0 &  5    3 &   61 &    11.026676\\
  b  a & 0    1 &  1   -1 &  0    0 &  0    2 &   61 &    56.314327\\[0.5ex]

  a  a & 0    1 &  1   -1 &  0    0 &  5    1 &   64 &    13.432528\\[0.5ex]

  a  b & 0    1 &  1   -1 &  0    0 &  1    5 &   76 &    18.376030\\
  b  a &-1    3 &  3   -4 &  0    0 &  3   -2 &   76 &    54.797520\\[0.5ex]

  a  b & 0    1 &  1   -1 &  0    0 &  5   -3 &   76 &    23.752563\\
  b  a & 0    1 &  1   -1 &  0    0 &  7   -9 &   76 &    49.421051\\[0.5ex]

  a  b & 0    1 &  1   -1 &  0    0 &  5   -1 &   79 &    17.171223\\
  b  a & 0    1 &  1   -1 &  0    0 &  2   -1 &   79 &    51.163399\\[0.5ex]
\end{tabular}
\end{center}
\end{table}

\begin{table}
\caption{Representatives for coincidence rotations with $\Sigma <
100$ for the icosahedral module.~II. Same as Tab.\ \ref{tab2a} for
  rotations along 2-fold symmetry axis.
\label{tab2b}}
\vspace{0.5cm}
\begin{center}
\begin{tabular}{c|c|c|c|c|c|c}
conjugates & $\kappa$ & $\lambda$ & $\mu$ & $\nu$ & $\Sigma$ & angle \\
\hline 
  a  a & 0    1 &  0    0 &  0    0 &  0    1 &    4 &    90.000000\\[0.5ex]

  a  a & 0    1 &  0    0 &  0    0 &  1    0 &    5 &    63.434948\\[0.5ex]

  a  a & 0    1 &  0    0 &  0    0 &  1    1 &    9 &    41.810310\\[0.5ex]

  a  a & 0    1 &  0    0 &  0    0 &  2    1 &   20 &    26.565052\\[0.5ex]

  a  a & 0    1 &  0    0 &  0    0 &  0    2 &   25 &    53.130100\\[0.5ex]

  a  b & 0    1 &  0    0 &  0    0 &  2    0 &   29 &    34.344078\\
  b  a & 0    1 &  0    0 &  0    0 &  2   -2 &   29 &    77.946892\\[0.5ex]

  a  a & 0    1 &  0    0 &  0    0 &  2   -1 &   36 &    48.189682\\[0.5ex]

  a  b & 0    1 &  0    0 &  0    0 &  1    2 &   41 &    30.900869\\
  b  a & 0    1 &  0    0 &  0    0 & -1    3 &   41 &    71.779510\\[0.5ex]

  a  b & 0    1 &  0    0 &  0    0 &  2    2 &   45 &    21.624634\\
  b  a &-4    8 &  0    0 &  0    0 &  0    2 &   45 &    74.754700\\[0.5ex]

  a  a & 0    1 &  0    0 &  0    0 &  3    2 &   49 &    16.601551\\[0.5ex]

  a  b & 0    1 &  0    0 &  0    0 &  3    1 &   61 &    19.387444\\
  b  a &-1    3 &  0    0 &  0    0 &  1    0 &   61 &    81.001404\\[0.5ex]

  a  a &-4    8 &  0    0 &  0    0 &  6   -8 &   81 &    83.620689\\[0.5ex]

  a  b & 0    1 &  0    0 &  0    0 &  4    2 &   89 &    13.463411\\
  b  a &-4    8 &  0    0 &  0    0 &  2    0 &   89 &    50.547329\\[0.5ex]

  a  a & 0    1 &  0    0 &  0    0 &  0    3 &  100 &    36.869896\\[0.5ex]
\end{tabular}
\end{center}
\end{table}

\begin{table}
\caption{Representatives for coincidence rotations with $\Sigma < 100$
  for the icosahedral module.~III. Same as Tab.\ \ref{tab2a} for
  rotations in mirror planes or general positions (last entry only).
\label{tab2c}}
\vspace{0.5cm}
\begin{center}
\begin{tabular}{c|c|c|c|c|c|c}
conjugates & $\kappa$ & $\lambda$ & $\mu$ & $\nu$ & $\Sigma$ & angle \\
\hline 
  a  b & 3   -4 &  1   -1 &  2    1 &  0    0 &   44 &    27.951245\\
  b  a &-3    5 &  0    1 &  1    2 &  0    0 &   44 &    31.212158\\[0.5ex]

  a  b &-1    3 &  2   -3 &  3    0 &  0    0 &   55 &    32.220027\\
  b  a & 3   -4 &  1   -1 &  4   -3 &  0    0 &   55 &    33.780077\\[0.5ex]

  a  b &-1    2 & -1    3 &  2    3 &  0    0 &   59 &    25.896548\\
  b  a & 0    1 &  3   -4 &  1    3 &  0    0 &   59 &    31.791262\\[0.5ex]

  a  a &-1    2 & -1    3 &  4   -1 &  0    0 &   64 &    29.364595\\[0.5ex]

  a  b & 2   -2 & -2    4 &  2    2 &  0    0 &   71 &    31.020216\\
  b  a & 1    0 & -3    5 &  4   -3 &  0    0 &   71 &    50.149411\\[0.5ex]

  a  b &-1    2 & -1    3 &  0    5 &  0    0 &   76 &    32.001147\\
  b  a &-2    4 & -2    4 &  0    4 &  0    0 &   76 &    30.228843\\[0.5ex]

  a  b &-3    5 &  0    1 &  3    0 &  0    0 &   79 &    23.521099\\
  b  a &-3    6 &  1   -1 &  2    1 &  0    0 &   79 &    34.169244\\[0.5ex]

  a  b &-1    2 & -1    3 &  4    1 &  0    0 &   80 &    21.724197\\
  b  a & 2   -2 & -2    4 &  4   -2 &  0    0 &   80 &    36.000000\\[0.5ex]

  a  a & 0    1 &  3   -4 &  3    1 &  0    0 &   81 &    25.322158\\[0.5ex]

  a  b &-1    3 &  2   -3 &  1    4 &  0    0 &   89 &    28.024012\\
  b  a & 2   -3 & -3    6 &  3   -1 &  0    0 &   89 &    33.772669\\[0.5ex]

  a  d &-1    3 &  2   -3 &  3    2 &  0    0 &   95 &    23.120394\\
  b  c &-1    3 &  2   -3 &  5   -4 &  0    0 &   95 &    37.840255\\
  c  b & 3   -4 &  1   -1 &  4   -1 &  0    0 &   95 &    21.809873\\
  d  a &-1    3 &  2   -3 & -3    8 &  0    0 &   95 &    48.040716\\[0.5ex]

  a  c &-2    4 & -2    4 &  2    2 &  0    0 &   99 &    23.316582\\
  b  d &-2    4 & -2    4 &  4   -2 &  0    0 &   99 &    27.162221\\
  c  a &-1    3 &  2   -3 & -1    6 &  0    0 &   99 &    35.483519\\
  d  b &-3    5 &  0    1 & -3    8 &  0    0 &   99 &    35.618152\\[0.5ex]

  a  b & 3   -4 &  1   -1 &  2    3 &  0    0 &  100 &    19.191145\\
  b  a & 3   -4 &  1   -1 & -2    7 &  0    0 &  100 &    31.294523\\[0.5ex]
\hline 
  a  b & 2   -2 &  2   -3 & -3    5 &  3    0 &  124 &    29.253916\\
  b  a & 0    1 &  5   -8 & -3    5 &  2    0 &  124 &    34.815460\\[0.5ex]
\end{tabular}
\end{center}
\end{table}

\begin{table}
\caption{Representatives for coincidence rotations with odd $\Sigma <
100$ for the cubic module $\ZZ[\tau]^{3}$.~I. The coordinates are given as
$m+\tau n$. The sixth column gives the orbit length. The last column
lists the disorientation angle.
\label{tab3a}}
\vspace{0.5cm}
\begin{center}
\begin{tabular}{c|r|r|r|r|r|c}
$\kappa$ & $\lambda$ & $\mu$ & $\nu$ & $\Sigma$ & orbit & angle \\
\hline

        1  0  &  0  0  &  0  0  &  0  0  &    1  &  1  &    0.00000\\[0.5ex]

        0  1  &  1  0  &  0  0  &  0  0  &    5  &  6  &    26.56504\\[0.5ex]

        1  1  &  1  0  &  0  0  &  0  0  &    9  &  6  &    41.81032\\
        1  0  &  1  0  &  1  0  &  0  0  &   9  &  4  &   60.00000\\[0.5ex]

        0  1  &  0  1  &  1  0  &  0  0  &   11  & 12  &   47.21220\\
        0  1  &  1  0  &  1  0  &  0  0  &   11  & 12  &   61.03971\\[0.5ex]

        0  1  &  0  1  &  0  1  &  1  0  &   19  &  8  &   20.72495\\
        0  1  &  1  0  &  1  0  &  1  0  &   19  &  8  &   26.10149\\
        1  1  &  1  1  &  1  0  &  0  0  &   19  & 12  &   30.22885\\
        1  1  &  1  0  &  1  0  &  0  0  &   19  & 12  &   56.75419\\[0.5ex]

       2  0  &  1  0  &  0  0  &  0  0  &   25  &  6  &   36.86989\\
       1  1  &  0  1  &  0  1  &  1  0  &   25  & 24  &   37.39882\\[0.5ex]

      -2  2  &  1  0  &  0  0  &  0  0  &   29  &  6  &   12.05311\\
       0  2  &  1  0  &  0  0  &  0  0  &   29  &  6  &   34.34407\\
       1  1  &  1  1  &  0  1  &  1  0  &   29  & 24  &   38.72486\\
       1  1  &  0  1  &  1  0  &  1  0  &   29  & 24  &   45.97971\\[0.5ex]

       1  2  &  1  1  &  1  0  &  0  0  &   31  & 24  &   34.81546\\
       1  1  &  1  1  &  1  1  &  1  0  &   31  &  8  &   35.12743\\
       1  2  &  0  1  &  1  0  &  0  0  &   31  & 24  &   48.36283\\
       1  1  &  1  0  &  1  0  &  1  0  &   31  &  8  &   53.02398\\[0.5ex]

       3 -1  &  1  0  &  0  0  &  0  0  &   41  &  6  &   18.22049\\
       1  2  &  1  2  &  1  0  &  0  0  &   41  & 12  &   18.95350\\
       2  1  &  1  0  &  0  0  &  0  0  &   41  &  6  &   30.90087\\
       1  2  &  1  0  &  1  0  &  0  0  &   41  & 12  &   36.92323\\
       2  0  &  0  1  &  1  0  &  0  0  &   41  & 24  &   44.08459\\
       0  2  &  0  1  &  1  0  &  0  0  &   41  & 24  &  47.75893\\[0.5ex]

       0  1  & -2  2  &  0  0  &  0  0  &   45  &  6  &   15.24529\\
       2  2  &  1  0  &  0  0  &  0  0  &   45  &  6  &   21.62464\\
       1  2  &  1  1  &  0  1  &  1  0  &   45  & 48  &   49.21481\\[0.5ex]

       2  3  &  1  0  &  0  0  &  0  0  &   49  &  6  &   16.60156\\
       0  2  &  1  1  &  1  0  &  0  0  &   49  & 24  &   29.53854\\
       2  0  &  1  0  &  1  0  &  1  0  &   49  &  8  &   38.21321\\
      -1  2  &  1  0  &  1  0  &  0  0  &   49  & 12  &   60.26300\\[0.5ex]

\end{tabular}
\end{center}
\end{table}

\begin{table}
\caption{Representatives for coincidence rotations with odd $\Sigma <
100$ for the cubic module $\ZZ[\tau]$.~II. Notation as in Tab.\ \ref{tab3a}.
\label{tab3b}}
\vspace{0.5cm}
\begin{center}
\begin{tabular}{c|r|r|r|r|r|c}
$\kappa$ & $\lambda$ & $\mu$ & $\nu$ & $\Sigma$ & orbit & angle \\
\hline
       2  0  &  0  1  &  1  0  &  1  0  &   55  & 24  &  33.78008\\
      -1  2  &  0  1  &  1  0  &  0  0  &   55  & 24  &   43.64693\\
       0  2  &  0  1  &  0  1  &  1  0  &   55  & 24  &   47.86958\\
       1  2  &  1  2  &  1  1  &  1  0  &   55  & 24  &  47.86958\\
       2  1  &  0  1  &  1  0  &  0  0  &   55  & 24  &   50.14141\\
       1  2  &  0  1  &  1  0  &  1  0  &   55  & 24  &   53.79745\\[0.5ex]

       2  1  &  1  0  &  1  0  &  0  0  &   59  & 12  &  42.69896\\
       3 -1  &  1  0  &  1  0  &  0  0  &   59  & 12  &   62.43954\\[0.5ex]
       2  0  &  0  1  &  0  1  &  1  0  &   59  & 24  &   25.89653\\
       1  2  &  1  2  &  0  1  &  1  0  &   59  & 24  &   35.23041\\
       0  2  &  0  1  &  1  0  &  1  0  &   59  & 24  &   55.84497\\
       1  2  &  1  1  &  1  0  &  1  0  &   59  & 24  &  41.14530\\

       0  1  &  3 -1  &  0  0  &  0  0  &   61  &  6  &   8.99856\\
      -2  2  &  1  0  &  1  0  &  1  0  &   61  &  8  &  11.02669\\
       1  3  &  1  0  &  0  0  &  0  0  &   61  &  6  &  19.38743\\
       1  1  &  2  0  &  1  0  &  0  0  &   61  & 24  &  36.95959\\
       2  2  &  0  1  &  1  0  &  0  0  &   61  & 24  &   39.92927\\
       2  2  &  1  1  &  1  0  &  0  0  &   61  & 24  &   41.49912\\
       0  1  & -2  2  &  1  0  &  0  0  &   61  & 24  &   54.33899\\
       0  2  &  1  0  &  1  0  &  1  0  &   61  &  8  &  56.31433\\[0.5ex]

       2  3  &  1  2  &  1  0  &  0  0  &   71  & 24  &   30.03838\\
       2  3  &  0  1  &  1  0  &  0  0  &   71  & 24  &   31.02022\\
      -1  2  &  0  1  &  1  0  &  1  0  &   71  & 24  &   38.59492\\
       0  2  &  1  1  &  1  1  &  1  0  &   71  & 24  &  38.59492\\
       0  2  &  1  1  &  1  0  &  1  0  &   71  & 24  &  39.35215\\
       2  1  &  0  1  &  0  1  &  1  0  &   71  & 24  &   53.42054\\[0.5ex]

       3 -1  &  1  0  &  1  0  &  1  0  &   79  &  8  &  17.17122\\
       0  1  & -2  2  &  1  0  &  1  0  &   79  & 24  &   23.52110\\
       1  1  & -1  2  &  1  0  &  0  0  &   79  & 24  &   33.58591\\
       1  3  &  1  1  &  1  0  &  0  0  &   79  & 24  &   45.25742\\
       1  1  &  2  0  &  1  0  &  1  0  &   79  & 24  &  45.27756\\
       2  2  &  0  1  &  0  1  &  1  0  &   79  & 24  &   50.99529\\
       2  1  &  1  0  &  1  0  &  1  0  &   79  &  8  &  51.16339\\
       2  2  &  1  1  &  1  1  &  1  0  &   79  & 24  &  55.79540\\[0.5ex]

\end{tabular}
\end{center}
\end{table}

\begin{table}
\caption{Representatives for coincidence rotations with odd $\Sigma <
100$ for the cubic module $\ZZ[\tau]$.~III. Notation as in Tab. \ref{tab3a}.
\label{tab3c}}
\vspace{0.5cm}
\begin{center}
\begin{tabular}{c|r|r|r|r|r|c}
$\kappa$ & $\lambda$ & $\mu$ & $\nu$ & $\Sigma$ & orbit & angle \\
\hline

       3 -1  & -2  2  &  0  0  &  0  0  &   81  &  6  &   6.37933\\
       2  0  &  2  0  &  1  0  &  0  0  &   81  & 12  &  38.94244\\
       2  1  &  1  1  &  0  1  &  1  0  &   81  & 48  &  48.42724\\
       2  3  &  1  1  &  1  1  &  1  0  &   81  & 24  &  58.45731\\[0.5ex]

       2  4  &  1  0  &  0  0  &  0  0  &   89  &  6  &   13.46340\\
       0  2  &  0  2  &  1  0  &  0  0  &   89  & 12  &   24.65171\\
       1  1  & -2  2  &  0  0  &  0  0  &   89  &  6  &    39.45264\\
       2  1  &  1  1  &  1  0  &  1  0  &   89  & 24  &  39.45264\\
       2  1  &  1  1  &  1  1  &  1  0  &   89  & 24  &  41.70632\\
       2  3  &  1  2  &  1  1  &  1  0  &   89  & 48  &  46.37142\\
       2  3  &  1  1  &  0  1  &  1  0  &   89  & 48  &   50.54736\\
      -2  2  & -2  2  &  1  0  &  0  0  &   89  & 12  &   59.54441\\[0.5ex]

       1  3  &  1  2  &  1  0  &  0  0  &   95  & 24  &  24.04593\\
       1  2  &  2  1  &  1  0  &  0  0  &   95  & 24  &  22.23439\\
       0  1  &  0  1  & -2  2  &  1  0  &   95  & 24  &  21.80987\\
       2  3  &  0  1  &  0  1  &  1  0  &   95  & 24  &  40.03725\\
       1  2  &  0  2  &  0  1  &  1  0  &   95  & 48  &  42.02099\\
       2  2  &  0  1  &  1  0  &  1  0  &   95  & 24  &  44.62797\\
       2  2  &  1  2  &  1  1  &  1  0  &   95  & 48  &  46.68387\\
       2  3  &  1  2  &  1  2  &  1  0  &   95  & 24  &  53.90720\\[0.5ex]

       2  3  &  2  3  &  1  0  &  0  0  &   99  & 12  &  11.78025\\
       2  3  &  1  0  &  1  0  &  0  0  &   99  & 12  &  23.31658\\
       1  3  &  1  0  &  1  0  &  0  0  &   99  & 12  &  27.16223\\
       2  2  &  1  2  &  0  1  &  1  0  &   99  & 48  &  33.71138\\
       1  1  &  1  1  &  2  0  &  1  0  &   99  & 24  &  35.61815\\
       1  1  & -1  2  &  1  0  &  1  0  &   99  & 24  &  45.51314\\
       2  2  &  1  1  &  1  0  &  1  0  &   99  & 24  &  45.51314\\
       1  2  &  0  2  &  1  1  &  1  0  &   99  & 48  &  45.93451\\
       0  1  &  0  1  &  3 -1  &  0  0  &   99  & 12  &  60.51462\\
       1  3  &  1  1  &  1  1  &  1  0  &   99  & 24  &  60.51462\\
\end{tabular}
\end{center}
\end{table}

\begin{table}
\caption{Representatives for coincidence rotations with even $\Sigma <
100$ for the cubic module $\ZZ[\tau]$.~IV. Notation as in Tabl.~\ref{tab3a}.
\label{tab3d}}
\vspace{0.5cm}
\begin{center}
\begin{tabular}{c|r|r|r|r|r|c}
$\kappa$ & $\lambda$ & $\mu$ & $\nu$ & $\Sigma$ & orbit & angle \\
\hline
        1  1  &  0  1  &  1  0  &  0  0  &    4 & 8  &  44.47752\\[0.5ex]

        2  1  &  1  1  &  1  0  &  0  0  &   16 & 8  &  31.04498\\
        2  3  & 1  1  &  1  0  &  0  0  &   16  &24  & 44.47752\\[0.5ex]

        1  3  & 0  1  &  1  0  &  0  0  &   20 & 24  & 36.00000\\
        0  1  &  3 -1  &  1  0  &  0  0  &   20 & 24  & 51.08879\\[0.5ex]

       2  3  &  1  3  &  1  0  &  0  0  &   36  & 8   & 15.52248\\
       0  3  &  1  1  &  1  0  &  0  0  &   36  & 24  & 38.99632\\
       1  1  &  0  1  & -2  2  &  1  0  &   36  & 24  & 41.81032\\[0.5ex]
 
       3  5  &  0  1  &  1  0  &  0  0  &   44  & 24  & 19.46459\\
       3  3  &  0  1  &  1  0  &  0  0  &   44  & 24  & 27.22766\\
       3  5  &  2  3  &  1  0  &  0  0  &   44  & 24  & 27.95126\\
       1  3  &  2  1  &  1  0  &  0  0  &   44  & 24  & 31.21215\\
       3  1  &  0  1  &  1  0  &  0  0  &   44  & 24  & 44.77236\\
      -3  3  &  0  1  &  1  0  &  0  0  &   44  & 24  & 44.87530\\
      -1  3  &  0  1  &  1  0  &  0  0  &   44  & 24  & 51.60709\\
      -1  2  &  0  1  &  3 -1  &  0  0  &   44  & 24  & 56.01215\\[0.5ex]

       3  3  &  2  3  &  1  0  &  0  0  &   64  &  8 & 13.43254\\
      -2  3  &  1  1  &  1  0  &  0  0  &   64  & 24  & 29.36458\\
       1  1  & -1  2  &  0  1  & -2  2  &   64  & 24  & 31.04498\\
       2  1  &  3 -1  &  1  0  &  0  0  &   64  & 24 & 50.48567\\
       1  3  &  2  0  &  0  1  &  1  0  &   64  & 48 & 50.48567\\[0.5ex]

       1  3  &  0  3  &  1  0  &  0  0  &   76  &  8 & 18.37603\\
       4  7  &  1  1  &  1  0  &  0  0  &   76  & 24 & 20.72495\\
       3  1  &  2  1  &  1  0  &  0  0  &   76  &  8 & 23.75257\\
       4  5  &  1  1  &  1  0  &  0  0  &   76  & 24 & 26.10149\\
       1  1  &  4 -1  &  1  0  &  0  0  &   76  & 24 & 32.00114\\
       3  1  &  0  2  &  0  1  &  1  0  &   76  & 48 & 42.11836\\
       2  1  &  1  1  & -2  2  &  1  0  &   76  & 48 & 43.06977\\
       4 -1  &  3 -1  &  1  0  &  0  0  &   76  &  8 & 49.42100\\
       1  2  &  0  1  &  3 -1  &  0  0  &   76  & 24 & 53.34293\\
      -1  3  &  2  0  &  0  1  &  1  0  &   76  & 48 & 53.34293\\
       3  0  &  0  1  &  3 -1  &  0  0  &   76  &  8 & 54.79753\\
       2  4  &  2  1  &  1  1  &  1  0  &   76  & 48 & 54.79753\\[0.5ex]

       1  2  &  1  1  &  0  1  & -2  2  &   80  & 48 & 51.08879\\
      -1  2  &  0  1  &  3 -1  & -2  2  &   80  & 48 & 26.56504\\
      -1  3  &  2  1  &  1  0  &  0  0  &   80  & 24 & 21.72421\\
       3  3  &  2  1  &  1  0  &  0  0  &   80  & 24 & 42.54018\\
      -2  3  &  3 -1  &  1  0  &  0  0  &   80  & 24 & 51.45900\\
       3 -1  &  6 -3  &  1  0  &  0  0  &   80  & 24 & 59.12300\\[0.5ex]

      4  7  &  3  5  &  1  0  &  0  0  &  100  & 24 & 19.19113\\
      2  5  &  1  3  &  1  0  &  0  0  &  100  & 24 & 31.29453\\
      3  0  &  1  1  &  0  1  & -2  2  &  100  & 24 & 37.39882\\
      3  5  &  0  2  &  0  1  &  1  0  &  100  & 48 & 37.39882\\
      2  5  &  2  2  &  1  1  &  1  0  &  100  & 48 & 44.47752\\
      0  4  &  1  1  &  0  1  &  1  0  &  100  & 48 & 53.13010\\
      5 -2  &  0  1  &  3 -1  &  0  0  &  100  & 24 & 60.18418\\

\end{tabular}
\end{center}
\end{table}

\end{document}